\documentstyle[12pt]{article}
\def\dalemb#1#2{{\vbox{\hrule height .#2pt
        \hbox{\vrule width.#2pt height#1pt \kern#1pt
                \vrule width.#2pt}
        \hrule height.#2pt}}}

\let\a=\alpha \let\b=\beta \let\g=\gamma \let\d=\delta \let\e=\epsilon
\let\z=\zeta  \let\q=\theta \let\i=\iota \let\k=\kappa
\let\l=\lambda \let\m=\mu \let\n=\nu \let\x=\xi \let\p=\pi \let\r=\rho
\let\s=\sigma \let\t=\tau \let\u=\upsilon \let\f=\phi \let\c=\chi 
\let\w=\omega      \let\G=\Gamma \let\D=\Delta \let\Q=\Theta \let\L=\Lambda
\let\X=\Xi \let\P=\Pi  \let\U=\Upsilon \let\F=\Phi

\def\nn{\nonumber} \def\bd{\begin{document}} \def\ed{\end{document}}
\def\ds{\documentstyle} \let\fr=\frac \let\bl=\bigl \let\br=\bigr
\let\Br=\Bigr \let\Bl=\Bigl 
\let\bm=\bibitem
\let\na=\nabla
\let\pa=\partial \let\ov=\overline
\def\ie{{\it i.e.\ }} 
\newcommand{\pr}{\paragraph{}}
\newcommand{\be}{\begin{equation}}
\newcommand{\ee}{\end{equation}}
\newcommand{\beba}{\begin{equation}\begin{array}{lcl}}
\newcommand{\eaee}{\end{array}\end{equation}}
\newcommand{\bea}{\begin{eqnarray}}
\newcommand{\eea}{\end{eqnarray}}
\newcommand{\ba}{\begin{array}}
\newcommand{\ea}{\end{array}}
\newcommand{\td}{\tilde}
\newcommand{\norsl}{\normalsize\sl}
\newcommand{\ns}{\normalsize}
\newcommand{\refs}[1]{(\ref{#1})}
\def\bal{{\mbox{\boldmath $\alpha$}}}
\def\bla{{\mbox{\boldmath $\lambda$}}}
\def\bbe{{\mbox{\boldmath $\beta$}}}
\def\bt{{\mbox{\boldmath $\tau$}}}
\def\bq{{\bf q}}
\def\bd{{\bf d}}
\def\bk{{\bf k}}
\def\bc{{\bf c}}
\def\bw{{\bf w}}
\def\bH{{\bf H}}
\def\bk{{\bf k}}
\def\bx{{\bf x}}
\def\boe{{\bf e}}
\def\a{\alpha}
\def\b{\beta}
\def\g{\gamma}
\def\c{\chi}
\def\d{\delta}
\def\e{\epsilon}
\def\ep{\varepsilon}
\def\FO{\phi}
\def\i{\iota}
\def\z{\psi}
\def\zb{\overline{\psi}}
\def\zt{\widetilde{\psi}}
\def\k{\kappa}
\def\l{\lambda}
\def\m{\mu}
\def\n{\nu}
\def\o{\omega}
\def\p{\pi}
\def\q{\theta}
\def\th{\theta}
\def\tc{\hat{\theta}}
\def\r{\rho}
\def\s{\sigma}
\def\st{\widetilde{\sigma}}
\def\sut{\utw{\sigma}}
\def\t{\tau}
\def\u{\upsilon}
\def\x{\xi}
\def\z{\zeta}
\def\w{\wedge}
\def\D{\Delta}
\def\F{\Phi}
\def\G{\Gamma}
\def\J{\Psi}
\def\L{\Lambda}
\def\O{\Omega}
\def\P{\Pi}
\def\Q{\Theta}
\def\U{\Upsilon}
\def\X{\Xi}
\def\f{\Phi_0}
\def\wtd{\widetilde}
\def\XH{\hat{X}}
\def\DH{\hat{D}}
\def\gh{\hat{g}}
\def\bh{\hat{b}}
\def\sg{\sqrt{-\g}}
\def\pa{\partial}
\def\gh{\hat{g}}
\def\bh{\hat{b}}
\def\mb{{\bar{m}}}
\def\nb{{\bar{n}}}
\def\rb{{\bar{r}}}
\def\sb{{\bar{s}}}
\def\ght{\hat{g}\kern-0.6em \widetilde{\raisebox{-0.12em}{\phantom{X}}}}
\def\bht{\hat{b}\kern-0.6em \widetilde{\raisebox{0.15em}{\phantom{X}}}}
\def\cF{{\cal F}}
\def\bG{{\bf G}}
\def\cL{{\cal L}}
\def\cM{{\cal M}}
\def\cG{{\cal G}}
\def\vf{\varphi}
\def\ft#1#2{{\textstyle{{\scriptstyle #1}\over {\scriptstyle #2}}}}
\def\fft#1#2{{#1 \over #2}}
\def\del{\partial}
\def\sst#1{{\scriptscriptstyle #1}}
\def\oneone{\rlap 1\mkern4mu{\rm l}}
\def\e7{E_{7(+7)}}
\def\td{\tilde}
\def\wtd{\widetilde}
\def\im{{\rm i}}
\def\bog{Bogomol'nyi\ }
\newcommand{\ho}[1]{$\, ^{#1}$}
\newcommand{\hoch}[1]{$\, ^{#1}$}
\newcommand{\ra}{\rightarrow}
\newcommand{\lra}{\longrightarrow}
\newcommand{\Lra}{\Leftrightarrow}
\newcommand{\ap}{\alpha^\prime}
\newcommand{\bp}{\tilde \beta^\prime}
\newcommand{\tr}{{\rm tr} }
\newcommand{\Tr}{{\rm Tr} } 

\begin{document}
\thispagestyle{empty}
\rightline{hep-th/9804096}
\vspace{1truecm}

\centerline{\bf \Large 
Cosmological Solution in M-theory on $S^1/Z_2$}

\vspace{1.2truecm}
\centerline{\bf Karim Benakli\footnote{e-mail: 
 karim@chaos.tamu.edu}}
\vspace{.5truecm}
{\em
\centerline{ Center for Theoretical Physics,
Texas A\&M University, College Station, Texas 77843
}}

\vspace{2.2truecm}


\vspace{.5truecm}

\begin{abstract}

We provide the first example of a cosmological solution  of  the
Horava-Witten supergravity. This solution is obtained by exchanging
the  role of time with the radial coordinate of the transverse space
to the five-brane soliton. On the boundary this corresponds to
rotating an instanton solution into a tunneling process in a space
with Lorentzian signature, leading to an expanding universe. Due to
the freedom to choose different  non-trivial  Yang-Mills backgrounds
on the boundaries, the two walls of the universe ( visible and hidden
worlds) expand differently. However at late times the anisotropy is
washed away by gravitational interactions.

\end{abstract}

\let\LARGE=\large \let\Large=\large \let\large=\normalsize

\vfill {\small

}

\newpage


Recent years have lead to  surprising new developments in our attempt
to build a fundamental theory. In particular it has been conjectured
that some (yet unknown) $M$-theory unifies all the known vacuua of
string theory as well as eleven dimensional supergravity. These vacuua
are believed to be interconnected through a network of dualities. Many
pieces of evidence have been gathered in this direction.

One of the simplest vacuua to analyze,  
for providing new phenomenological or cosmological scenarios, is the strongly 
coupled limit of heterotic string theory. Horava and Witten have identified 
it with a $Z_2$ orbifold \cite{hw}. In the infrared limit, it has 
the description of an eleven dimensional supergravity on a manifold with 
boundaries. Gravitational anomalies are canceled by coupling 
the supergravity in each boundary with  an $E_8$ super--Yang--Mills theory.

Unlike usual orbifold compactifications of string theory, in the Horava-Witten orbifold 
the communication between the two boundaries is due only to
gravitational interactions. There is no gauge bosons propagating in
the bulk. This leads to the picture of two parallel worlds (the
observable and the hidden world) communicating through gravity.

It has been pointed out recently that cosmological solutions can be
generated  from p-brane solutions [2--4]. The process consists in
the  exchange the role of  time with the radial coordinate of the
transverse space\cite{bfs}. In this note we will use a similar approach to 
obtain a cosmological solution to  supergravity
theory on an eleven dimensional manifold with boundaries.  A large
number of cosmological models have been derived in the past from a
variety of vacuua of  M-theory. These include solutions of type II
strings with both R-R and NS-NS background fields as well as 
weakly coupled heterotic string compactifications [2--21].  We
will promote the discussion to the level of the strongly coupled
limit of heterotic string theory.

In Horava-Witten model the eleven dimensional bulk theory is described 
by a graviton supermultiplet. Its degrees of freedom are the metric $g_{IJ}$, 
the three--form
gauge potential $C_{IJK}$ with the associated field strength components
$G_{IJKL}$  and the gravitino.  Here $I,J$... take values $0,1,...,9,11$.
In addition on each boundary lives a super--Yang--Mills
ten--dimensional supermultiplet with a gauge field  ${\cal A}^a_A$ and the
corresponding  gauginos ($A,B...=0,1,...,9$). The gauge field strength 
is denoted by $F^{a(1,2)}_{AB}$. The bosonic part of the
corresponding  lagrangian (up to two  derivatives) takes the form:

\beba 
S &=& \int d^{11}x  \frac{1}{2\k^2}\sqrt{-g}
 \left[{-R-{\frac{1}{24}} G_{IJKL}G^{IJKL} }\right] \nn\\ 
&&-\int d^{11}x \frac{1}{\k^2}\frac{\sqrt{2}}{3456}\epsilon^{I_1...I_{11}}
 C_{I_1I_2I_3}G_{I_4...I_7}G_{I_8...I_{11}}\nn\\
 && -\int d^{11}x
 \delta(x^{11})\frac{1}{4\l^2}\sqrt{-g_{10}}\ {\rm tr}(F^{(1)}_{AB}
 F^{(1)AB}) \nn\\
 && - \int d^{11}x \delta (x^{11}-\pi \r )\frac{1}{4\l^2}
 \sqrt{-g_{10}}\  {\rm tr}(F^{(2)}_{AB}F^{(2)AB})  
\label{LS}
\eaee where $\l^2 = 4\pi (4\pi\k^2)^{2/3}$ [1,24]. The orbifold
segment is parametrized by  $x^{11}\in [0 ,\pi\r ]$. The metric $g_{10}$ is the
restriction of the eleven dimensional metric to the boundaries.

The Einstein equation are given by 
\beba 
R_{MN}-\frac{1}{2}g_{MN}R &=&
- \frac{1}{6}\left( G_{MJKL}{G_N}^{JKL} -\frac{1}{8}
g_{MN}G_{IJKL}G^{IJKL}\right)\nn\\ 
&& -\d (x^{11})\frac{\k^2}{\l^2}
(g_{11,11})^{-\frac{1}{2}}\left[  {\rm tr}(F^{(1)}_{AC}{F^{(1)}_B}^C)
-\frac{1}{4}g_{AB}{\rm tr}(F^{(1)}_{CD}F^{(1)CD})\right]\nn\\ 
&& -\d(x^{11}-\pi\r)\frac{\k^2}{\l^2} (g_{11,11})^{-\frac{1}{2}}\left[  {\rm
tr}(F^{(2)}_{AC}{F^{(2)}_B}^C) -\frac{1}{4}g_{AB}{\rm
tr}(F^{(2)}_{CD}F^{(2)CD})\right]\; . 
\label{Einst} 
\eaee

The  equations of motion of the three-form $C_{IJK}$ take the form:
\bea 
\partial_M\left(\sqrt{-g}G^{MNKL}\right)&=&\frac{\sqrt{2}}{1152}
\epsilon^{NKLI_4...I_{11}}\, G_{I_4...I_7}G_{I_8...I_{11}}\nn\\
&-&\frac{\k^2}{\l^2}\frac{1}{432}\d
(x^{11})\epsilon^{11NKLJ_5...J_{11}} \o_{J_5J_6J_7}G_{J_8...J_{11}} \; ,
\label{eomG} 
\eea 
where $\o_{J_5J_6J_7}$ is the Yang--Mills
Chern--Simons form.  The four-form  field strength $G_{ABCD}$ is also
subject to the Bianchi identity~\cite{hw}: 
\be 
(dG)_{11ABCD} =
-3\sqrt{2}\frac{\k^2}{\l^2}\d (x^{11})\left( {\rm
tr}(F_{[AB}F_{CD]})-\frac{1}{2} {\rm tr}(R_{[AB}R_{CD]})\right)\; .
\label{Bianchi} 
\ee

The boundary gauge fields satisfy 
\be 
D_AF^{iAB}=\frac{1}{\sqrt{2}}
(g_{11,11})^{\frac{1}{2}} F_{DE}^i  G^{11BDE}
+\frac{1}{1152}\frac{1}{\sqrt{-g_{10}}}\, \epsilon^{BI_2...I_{10}}
{\cal A}^i_{I_2} G_{I_3...I_6}G_{I_7...I_{10}}\; , 
\label{eqF} 
\ee 
with $D_A$ the space--time and gauge covariant derivative.

We want to exhibit a solution with a metric in eleven dimension of the
form (in polar coordinates):
\begin{equation}
ds^2 = e^{2A}\left( -q(dx^0)^2+dx^\m dx^\n\d_{\m\n}\right) +
e^{2B}\left( \frac{d\t^2}{q} + \t^2 d\Omega^2 _{3,k} \right) +  e^{2B}
(dx^{11})^2,
\label{newmet}
\end{equation}
where $q$ will later be taken equal to $-1$ later. The unit spherical
 volume element is given by:
\begin{equation}
d\Omega_{3,k}^2 = d\chi ^2 +\frac{\sin^2\! \left( \sqrt{k}
\chi\right)}{k} \left( d\theta ^2 + \sin^2 \!\theta d\omega^2\right).
\end{equation}
 
Taking $k=q$ the metric (\ref{newmet}) can be transformed to:  
\beba
ds^2 &=& e^{2A}\left( -dt^2+ dx^\m dx^\n\d_{\m\n}\right) + \nn\\ &&
e^{2B}\left( d r^2 + r^2 \left( d\phi ^2 +\sin^2\! \left( \phi \right)
\left( d\theta ^2 + \sin^2 \!\theta d\omega^2\right) \right) \right)
\nn\\ &&+  e^{2B} (dx^{11})^2,
\label{oldm}
\eaee through the changement of variables:

\be t=\sqrt{q}x^0 \, \ \ \ \ \, \phi= \sqrt{q} \chi \ \ \ \ \,
r=\frac{\t}{ \sqrt{q}}. 
\label{varc}
\ee 

We can now use the fact that if  some function $A$ and $B$, as well as a background configuration of gauge fields \ref{oldm} solves  the
equations of motion and satisfies all the constraints, then 
the new metric  (\ref{newmet}) obtained through a
simple coordinate transformation will also do.   A
particular non trivial solution with a metric of the form (\ref{oldm})
is the five-brane  soliton. This solution was obtained in  the weakly
coupled regime by \cite{str} and elevated to a solution of  the
strongly coupled theory by \cite{w}, then successively revisited by \cite{ovr} and \cite{dud}. Here we will closely follow the discussion of \cite{ovr}. For the five-brane the functions $A$, $B$ and the
four-form $G$ are  related by $A=-B/2=-C/6$ and  
\be 
G_{mnrs} =
\pm\frac{1}{\sqrt{2}}e^{-\frac{8C}{3}}{\epsilon_{mnrs}}^t\partial_t
e^C\;
\label{ffans}
\ee 
with $m,n,...=6,...,9,11$. The function $C$
depends only on  $x^{11}$ and $r$.  We first solve the equations of motion in
the metric  we will extract a cosmological
solution by  exchanging the role of time with the radial coordinate of
the transverse space through (\ref{varc}).

In our ansatz (\ref{ffans}) the only non vanishing components of  the
four-form field strength are $G_{abcd}$ and $G_{11abc}$ with
$a,b,...=6,...,9$.  For such configuration $G\wedge G$ vanishes  and
the equation of motion becomes $d * G= 0$.  In (\ref{ffans}) $*G$ is
a closed form so the equation of motion are automatically satisfied.

The Bianchi identity involves the gauge fields at the boundaries that
need to  be specified.  A non-trivial Yang-Mills background is obtained 
with  a field strength
self--dual with respect to the metric: 
\begin{equation}
ds_4 ^2 =  \frac{d\t^2}{q} + \t^2 d\Omega_{3,k}^2 .
\end{equation}
We choose a configuration which in the case $q=1$
corresponds to an  $SU(2)$ instanton. For $q=-1$, the gauge vectors
take value in a subset of  $SL(2,C)$ and the background considered can
be interpreted as mapping of a pseudo-sphere in the target space on a
a pseudo-sphere of the gauge group space \cite{bfs}.

It is useful to present the expressions for the gauge fields for a generic 
value of $q$ as they were derived in \cite{bfs}.  The gauge vector potential takes the form:

\begin{equation} \label{grupp}
{\cal A}\left( r\right) = \gamma\left( \t\right) \sqrt{q}g^{-1} dg,
\end{equation}
$g^{-1} dg$ is the Cartan-Maurer form where
\begin{equation} \begin{array}{ll}
g = &  \cos \sqrt{q} \chi - i \cos \omega \sin \theta \sin \sqrt{q}
\chi  \sigma_1 - i \sin \omega \sin \theta \sin \sqrt{q} \chi
\sigma_2\\ & -i \cos \theta \sin \sqrt{q} \chi \sigma_3 \,
, \end{array}
\end{equation}
with
  \begin{equation} 
\gamma\left( r\right) = \frac{
 \frac{\t^2}{\sqrt{q}\s^2 } } {\sqrt{q} \frac{\t^2}{\sqrt{q} \s^2} +1} ,
\end{equation}
where $\sqrt{q}\s^2 $ is an integration constant. The associated field
strength is given by: 
\be F^a = \fft{2\sqrt{q} \s^2}{(\t^2 + \s^2)^2} \eta^a_{mn}
dy^m\wedge dy^n\ , 
\ee 
where $a$ is an adjoint index of $SU(2)$ for
$q=1$ or its analytical continuation in $SL(2,C)$ for $q=-1$,
$\eta^a_{mn}$ are the corresponding  't Hooft symbols. Thus,  
\be 
\mbox{tr}(F_{[AB}F_{CD]})= -\sqrt{q} \epsilon_{ABCD}\fft{16 q \s^4}{(\s^2 +\t^2)^4}\; . 
\label{trf2} 
\ee

Following \cite{ovr}, we denote $e^C$ by $\Phi$ and write the solution
as : \be \Phi = \Phi_0+\phi, \label{ansa} \ee where $\Phi_0$ 
is a function of $r$
only, while $\phi$ is a function of both $r$ and $x^{11}$ with
vanishing average on $x^{11}$. $\phi$ contains the
corrections of the strong coupling limit of the heterotic string
theory.  The Bianchi identities can be written as an homogeneous
condition  $dG=0$ in the bulk  supplemented with the boundary
conditions \footnote{ Note that after wick rotation these quantities become imaginary, as expected because they represent a tunneling effect.}:

\bea \left. G_{ABCD}\right|_{x^{11}=0} =  -\frac{3}{\sqrt{2}}
 \frac{\k^2}{\l^2} \mbox{tr}(F^{(1)}_{[AB}F^{(1)}_{CD]})\nn\\
 \left. G_{ABCD}\right|_{x^{11}=\pi\r} =  +\frac{3}{\sqrt{2}}
 \frac{\k^2}{\l^2} \mbox{tr}(F^{(2)}_{[AB}F^{(2)}_{CD]})\; .
 \label{Bianchi3} \eea

They lead to the differential equations (we assume $q=\pm1$):

\be
 \frac{1}{r^3}\partial_r
\left(r^3 \partial_r \Phi_0\right) =\frac{q \k^2}{4 \pi \r \l^2} 
\left(\fft{16 \s_1^4}{(\frac {\s_1^2}{q} +r^2)^4}+\fft{16 \s_2^4}
{(\frac {\s_2^2}{q} +r^2)^4}\right)
\label{Fi0}
\ee

and 
\be 
\frac{1}{r^3}\partial_r \left(r^3 \partial_r \phi \right) +
\partial_{11}^2\phi=0
\label{fi} 
\ee

The function $\phi$ is subject to the boundary conditions: \be
\partial_{11} \phi |_{x^{11}=0} = -q\frac{\k^2}{4\l^2}\fft{16 \s_1^4}{(
\frac {\s_1^2}{q} + r^2 )^4}, \;\quad
\partial_{11} \phi |_{x^{11}=\pi\r} = q\frac{\k^2}{4\l^2}\fft{16 \s_2^4}
{(\frac {\s_2^2}{q} +r^2)^4}\;
.\label{bound0} 
\ee

We do not reproduce here the different steps of the solution and refer
for details to the original work by \cite{ovr}. After making the
changement of variables  (\ref{varc}), we obtain:

\be 
\Phi_o = 1 + \frac{2 \k^2}{\pi\r \lambda^2} \left ( \frac{2
\sigma_1^2+ \t^2}{(\sigma_1^2 + \t^2)^2} + \frac{2 \sigma_2^2+
\t^2}{(\sigma_2^2 + \t^2)^2} \right ) \; . 
\label{fio}
\ee 
and $\phi$ is given as an expansion in  $(\pi\r)^2/\s^2$: 
\be
\phi = \sum_{n=0}^{n=\infty} \phi_n\label{phi}
\ee
 The first terms of the expansion are given by:

\be \phi_0 = 48 \frac{\k^2}{\lambda^2} \left ( P_0 (x^{11})
\frac{\sigma_1^4} {(\sigma_1^2 + \t^2)^4}+  Q_0 (x^{11})
\frac{\sigma_2^4} {(\sigma_2^2 + \t^2)^4} \right )  \ee 
and  
\be 
\phi_1
= 768\; \frac{\k^2}{\lambda^2} \left ( P_1 (x^{11}) \frac{3 \t^2 -2
\sigma_1^2} {(\sigma_1^2 + \t^2)^6}+  Q_1 (x^{11}) \frac{3 \t^2 -2
\sigma_2^2} {(\sigma_2^2 + \t^2)^6} \right ) \; ,
\label{result}
  \ee

with:

\be P_0 = - \frac{(x^{11})^2}{4 \pi\r} + \frac{x^{11}}{2} - \frac{\pi
 \r}{6}\; , \quad Q_0 = - \frac{(x^{11})^2}{4 \pi\r} + \frac{\pi
 \r}{12} 
\label{eq:pzero}
\ee and  \be P_1 = \frac{(x^{11})^4}{48 \pi\r} -
\frac{(x^{11})^3}{12}+ \frac{\pi\r (x^{11})^2}{12} -
\frac{\pi^3\r^3}{90}, \; Q_1= \frac{(x^{11})^4}{48 \pi\r} -
\frac{\pi\r (x^{11})^2}{24} + \frac{7 \pi^3\r^3}{720}
\label{exp}
\ee

We thus found a cosmological solution of the form:

\begin{equation}
ds^2 = \Phi^{-1/3}\left( (dx^0)^2+dx^\m dx^\n\d_{\m\n}\right) +
\Phi^{2/3}\left( d\t^2 + \t^2 d\Omega^2 _{3,-1} \right) +  
\Phi^{2/3} (dx^{11})^2,
\label{finmet}
\end{equation}
with $\Phi$ given by (\ref{fio}) to (\ref{exp}). To have a time
coordinate which goes from $-\infty$  to $+\infty$ one makes the
rescaling  $\tau=e^{\eta}$. The choice $\tau=e^{-\eta}$ lead to a
similar branch  disconnected from this one.

We have thus transformed the gauge five-brane solution (in the form
given by \cite{ovr}) to a non-singular cosmological solution\footnote
{Extending it to the case of neutral five-brane \cite{duff} is a
trivial issue and leads to a singular cosmology.}. W e would like to
discuss now some interesting features of our solution. A four
dimensional solution is obtained by considering the six-dimensional transverse
space parametrized by $x^0,...,x^5$ as compact.

For $t\to 0$ or equivalently $\eta\to -\infty$ the size of the compact
internal space is finite. However the radius of the pseudo-sphere of
the universe goes to zero. Through tunneling effect a bubble with a
new expectation value of the field $\Phi$  (the  dual of the four-form
$G$) is created at $\eta\to -\infty$ and expands. 
This event is at finite proper time 
from any point in
the future. The initial conditions on the two boundaries may be
different ( if $\s_1 \neq \s_2$). Thus the tunneling on the two
boundaries follows different time dependence. On the $x^{11}=0$
boundary :

\be
\Phi= 1 + \frac{2 \k^2}{\pi\r \lambda^2} \left ( \frac{2
\sigma_1^2+ \t^2}{(\sigma_1^2 + \t^2)^2} + \frac{2 \sigma_2^2+
\t^2}{(\sigma_2^2 + \t^2)^2} \right )+  8 \frac{\k^2\pi \r}{\lambda^2} 
\left ( 
- \frac{\sigma_1^4} {(\sigma_1^2 + \t^2)^4}+  
\frac{\sigma_2^4} {2(\sigma_2^2 + \t^2)^4} \right )+... \; ,
\label{edge1}
 \ee  

while on the other edge $x^{11}=\pi \r$:
\be
\Phi= 1 + \frac{2 \k^2}{\pi\r \lambda^2} \left ( \frac{2
\sigma_1^2+ \t^2}{(\sigma_1^2 + \t^2)^2} + \frac{2 \sigma_2^2+
\t^2}{(\sigma_2^2 + \t^2)^2} \right )+  8 \frac{\k^2\pi \r}{\lambda^2} 
\left ( 
\frac{\sigma_1^4} {2 (\sigma_1^2 + \t^2)^4}-  
\frac{\sigma_2^4} {(\sigma_2^2 + \t^2)^4} \right )+... \; ,
\label{edge2}
 \ee  
where the dots are for higher orders of the expansion in $(\pi\r)^2/\s^2$.

At later times as $t\to +\infty$ or equivalently $\eta\to +\infty$, we
find $\Phi \to 1$. Thus the solution describes an expanding four
dimensional flat universe. The expansion anisotropy between the two
boundaries is washed away by gravitational interactions between
them. It is also interesting to notice that the volume of the internal
space and the size of the segment remain finite (small) and thus we
have a mechanism of spontaneous compactification.

It is tempting to find a relation between our toy example and the
recent work  of Hawking and Turok \cite{haw}. Our field $\Phi$
(obtained through dualization of the four form) or more precisely
$\Phi_0$ would be identified with the scalar field considered in
\cite{haw}.

We have  illustrated the possibility of having different initial
 conditions on the two boundaries of the universe which might lead to
 different expansions at early time. This might also have implication for 
dark matter physics at early times if as suggested in \cite{ben} 
it lives on the hidden wall of the universe.

I wish to thank Michael Duff, Jian Xin Lu, Dimitri Nanopoulos and 
particularly Chris Pope for useful discussions. I also thank  
Per Sundell and Tran Tuan for help with technical issues during the early 
stages of this work. This work was supported  by DOE grant 
DE-FG03-95-ER-40917. 


\end{document}